\documentclass[numbers,webpdf,imaiai]{ima-authoring-template}%

\graphicspath{{Fig/}}

\usepackage{natbib} 
\usepackage{soul}
\theoremstyle{thmstyletwo}%
\numberwithin{equation}{section}

\begin{document}
\bibliographystyle{apalike}
\setcitestyle{authoryear}
\DOI{DOI HERE}
\copyrightyear{2024}
\vol{00}
\pubyear{2024}
\access{Advance Access Publication Date: Day Month Year}
\appnotes{Paper}
\copyrightstatement{Published by Oxford University Press on behalf of the Institute of Mathematics and its Applications. All rights reserved.}
\firstpage{1}


\title[Levelling Up Learning]{Exploring the impact of gamification on engagement in a statistics classroom}

\author{Eilidh Jack* \ORCID{0000-0002-0017-8242} AND Craig Alexander\ORCID{0000-0001-6734-747X}
\address{\orgdiv{School of Mathematics and Statistics}, \orgname{University of Glasgow}, \orgaddress{\street{Glasgow}, \postcode{G12 8QQ}, \country{United Kingdom}}}}
\author{Elinor M Jones \ORCID{0000-0003-2551-8765}
\address{\orgdiv{Department of Statistical Science}, \orgname{University College London}, \orgaddress{\street{London}, \postcode{WC1E 7HB}, \country{United Kingdom}}}}

\authormark{Jack et al.}

\corresp[*]{Corresponding author: \href{email:Eilidh.Jack@glasgow.ac.uk}{Eilidh.Jack@glasgow.ac.uk}}

\received{Date}{0}{Year}
\revised{Date}{0}{Year}
\accepted{Date}{0}{Year}


\abstract{In recent years, the integration of gamification into educational settings has garnered significant attention as a means to enhance student engagement and learning outcomes. By leveraging gamified elements such as points and leaderboards, educators aim to promote active participation, motivation, and deeper understanding among students. This study investigates the effect of gamification on student engagement in a flipped statistics classroom environment. The findings suggest that gamification strategies, when effectively implemented, can have a positive impact on student motivation and engagement. This paper concludes with recommendations for educators, potential challenges such as superficial engagement and demotivation, and future directions for research to address these challenges and further explore the potential of gamification in fostering student success.}
\keywords{Gamification; Flipped learning; Student engagement; Active learning.}


\maketitle

\section{Introduction}\label{secintro}

Over recent years, the rapid advancement in teaching technology has led to an increased interest in the adoption of active learning approaches. Active learning shifts instruction to a learner-centred approach and incorporates a wide range of instructional strategies which may include student interaction, collaboration, problem solving, or reflection time, among many others (\cite{Hartikainen19}). Specifically in higher education, institutions are evolving their teaching away from a focus on the teacher, to a more distributed learning approach where students can study at a time and pace that suits them. Many have developed their own Learning and Teaching Strategy placing emphasis on such approaches, such as at the lead authors’ institution, The University of Glasgow\footnote{\url{https://www.gla.ac.uk/media/Media_842671_smxx.pdf}}.

A common approach which allows for the implementation of active learning is through flipped learning. Flipping the classroom is enabled through information and content instruction taking place outside of the classroom with in-class time instead being used for consolidation of learning, for example through interacting with their peers, solving problems and applying conceptual knowledge (\cite{Cronhjort17}). The flipped classroom thus facilitates active learning by requiring students to engage with the pre-class content, which in turn frees up class time for interactive activities that would not otherwise be possible (\cite{Wood23}).

There is growing evidence that a flipped learning approach can – when carefully implemented – benefit students. The meta analyses by \cite{Lo2017} and \cite{Bredow21} point to increased learning in the mathematical sciences when a flipped model is used in place of traditional lectures. There is also evidence that this mode of learning boosts student enjoyment, for example in terms of perceived understanding (e.g. \cite{Touchton15}). More generally, flipped learning can be an effective strategy for dealing with large lecture courses where student engagement and interaction would otherwise be infrequent (\cite{Marcey14}).

Despite the positive evidence in favour of a flipped classroom, the benefits can only be realised when students are motivated to complete out-of-class activities (\cite{Huang18}). Since the change in teaching style places responsibility on the students to take ownership of their learning, this can lead to issues with student engagement with pre-lecture activities including possible resistance from students (\cite{Stone12}).

A sub-optimal strategy to ensure that students keep on track is to use some of the in-class time to summarise the material which students should have studied in advance (\cite{Wilson13}). This reduces time available for consolidating understanding and runs the risk of a vicious cycle where students learn to rely on these summaries rather than complete the work before the session. Alternatively, adaptations can be made to a course curriculum to encourage engagement, for example using frequent summative and formative assessment after lessons (\cite{Owens20}).

A further option, which does not require changes to course curriculum, is to embed a ‘gamification’ strategy on the virtual learning environment that hosts the learning material (\cite{Ekici21}). Gamification aims to utilise elements of classic game design into a non-game context (\cite{Deterding11}) and awards points for participation in certain activities. In a virtual learning environment, course leaders can specify additional constraints on rewards such as timeliness of engagement and/or minimum level of interaction with - or achievement on - an activity. Strategies that reward active participation with the pre-lecture work have been shown to enhance student motivation to participate in the learning process and complete tasks prior to lecture sessions in flipped learning classrooms (\cite{Lo17}, \cite{Buckley16}, \cite{Dahlstrom12}).

The purpose of this study is to investigate the effects of introducing a gamification element to flipped learning materials in a mathematical sciences course. The study will focus on levels of engagement with such materials before and after the gamification intervention is introduced.  Section 2 describes the nature of the courses under consideration, while section 3 outlines the motivation for implementing a gamification strategy in these courses and how this was achieved. In section 4, we report on the remarkable increase in student engagement with pre-lecture materials after a game-based approach was employed. We conclude in section 5 with a discussion of the potential of gamification along with its potential drawbacks.

\section{Course information}\label{sec2}

At the University of Glasgow, the School of Mathematics and Statistics offer courses in introductory Statistics to undergraduate students across the University. These level 1 courses are typically taken in year 1 or 2 (pre-honours) of a student's university career, which typically lasts 4 years. In year 1, single honours students will take courses in three subjects, one of which will be the subject they have chosen to specialise in. Any of these three courses can be considered in year 2, and students have the option to switch their chosen degree programme to any of these subjects conditional on meeting the requirements for progression. The purpose of the level 1 statistics courses is predominately to prepare students who are interested in studying statistics at higher levels and, given that most students will have little experience of statistics when they arrive at university, introduce students to the subject with the hope that students will choose to continue studying statistics in future years. There are two level 1 courses taught over two semesters (semester 1: Sep-Dec and semester 2: Jan-Mar) with final exams taking place at the end of each semester. Given the growing awareness of the importance of statistics across many disciplines, the courses have become increasingly popular over the past few years with 200+ students enrolled.

To align with the University of Glasgow Learning and Teaching strategy, the courses were re-designed for 2020-21 with the aim of evolving the approach to learning and teaching towards student-centered, active learning. The approach taken for these courses was to introduce a flipped learning model, with a variety of online learning materials for students to complete before live teaching sessions. In-person sessions were then used to consolidate learning through  a variety of interactive activities for students to complete individually or in small groups. For the purpose of this study we will call these online learning materials, `pre-lecture activities'.

\subsection{Data}\label{subsecdata}
The data for this study are taken from Moodle activity logs. These can be used to view all student interactions with Moodle, including engagement with the activities of interest for the courses under consideration here. Data such as these are becoming increasingly available and the use of data in the support of student learning is commonly referred to as learning analytics (\cite{Leitner17}). The research ethics associated with this study required students to consent to their data being used as part of this project.\footnote{At this institution, staff may undertake learning analytics activities which enable interventions to assist students through the analysis of their own data. However, when such projects are intended for external dissemination they also require ethical approval. This study falls into the latter category and ethical approval was sought and granted in advance of this study taking place (application number 300220006).} Consent was requested during live teaching sessions and was granted for $n=100$ (out of a total of 257) students in 2022-23 and $n=77$ (out of a total of 280) students in 2023-24. Unfortunately we were unable to obtain retrospective ethical approval to access data from 2020-21 (when flipped learning was introduced) and 2021-22 due to the opt-in nature of the study. We can therefore  only provide results and comparisons for engagement using the 2022-23 and 2023-24 cohorts.

Students attended three in-person sessions a week, and before each of these were asked to complete the following pre-lecture activities: 
\begin{itemize}
    \item \textbf{Textbook readings} - students were asked to read a section of the course textbook in advance of the associated in-person session. The purpose of the readings was to introduce students to the course material. Completion of these can only be tracked through students manually marking an activity as complete via on Moodle. 
    \item \textbf{Formative reading quizzes} - students were asked to complete a short quiz  after reading  the corresponding textbook section. The purpose of the quizzes was for students to consolidate their learning and test their understanding of the course material in advance of the corresponding in-person session. Information on attempts for all quizzes is automatically stored in Moodle. It is therefore possible to determine which students have completed a quiz, when they started/finished the quiz, and their performance on the quiz. We can therefore track engagement in the quizzes and compare this with different cohorts as well as monitoring performance to inform the active learning sessions. These will be referred to as pre-lecture quizzes.
\end{itemize}

\section{Motivation and intervention}\label{secmotivation}

The pre-lecture activities were made available to students via the University of Glasgow's virtual learning environment, Moodle \footnote{\url{https://moodle.org/}}. As such, it is straightforward for course instructors to check if students are engaging with these activities and crucially, given the flipped learning approach, whether they are engaging with them in advance of the in-person sessions.

For the purpose of this study, only engagement with the formative reading quizzes (pre-lecture quizzes) is measured. This means that we cannot measure and compare engagement across all pre-lecture activities and we may be misrepresenting true engagement as some students may choose to read the textbook but not complete the reading quiz (or vice versa). However, this should give a more reliable estimate of meaningful engagement in activities since students have to engage somewhat with the activity (e.g. start and submit the quiz) rather than being able to manually mark that the activity is complete by checking a button. This does not fully guard against non-meaningful engagement (e.g. a student could begin the quiz, complete no questions, submit, and still be counted as having completed this activity) but performance in these activities (e.g. questions attempted and marks achieved) was also checked to understand the extent of meaningful engagement.

Despite students indicating that they enjoyed the flipped learning approach through course evaluation forms, when looking at the data it was clear that engagement in the pre-lecture quizzes was extremely low. For example, the average completion of pre-lecture quizzes in advance of the active learning session taking place was 6.88\% in semester 1 of 2022-23. This was of concern to course instructors given the in-class sessions involved spending less time presenting material and more time engaging in active learning activities as per the goal of flipped learning. This pattern was persistent over the academic years 2020-21, 2021-22 and semester 1 of 2022-23. Given that the introduction of the new courses coincided with the Covid-19 pandemic, it was also difficult to disentangle the lack of engagement with the flipped learning material, and the general lack of engagement HE institutions saw over the course of the pandemic (\cite{Wester21}). 

\begin{figure*}%
\centering
\includegraphics[width=\textwidth]{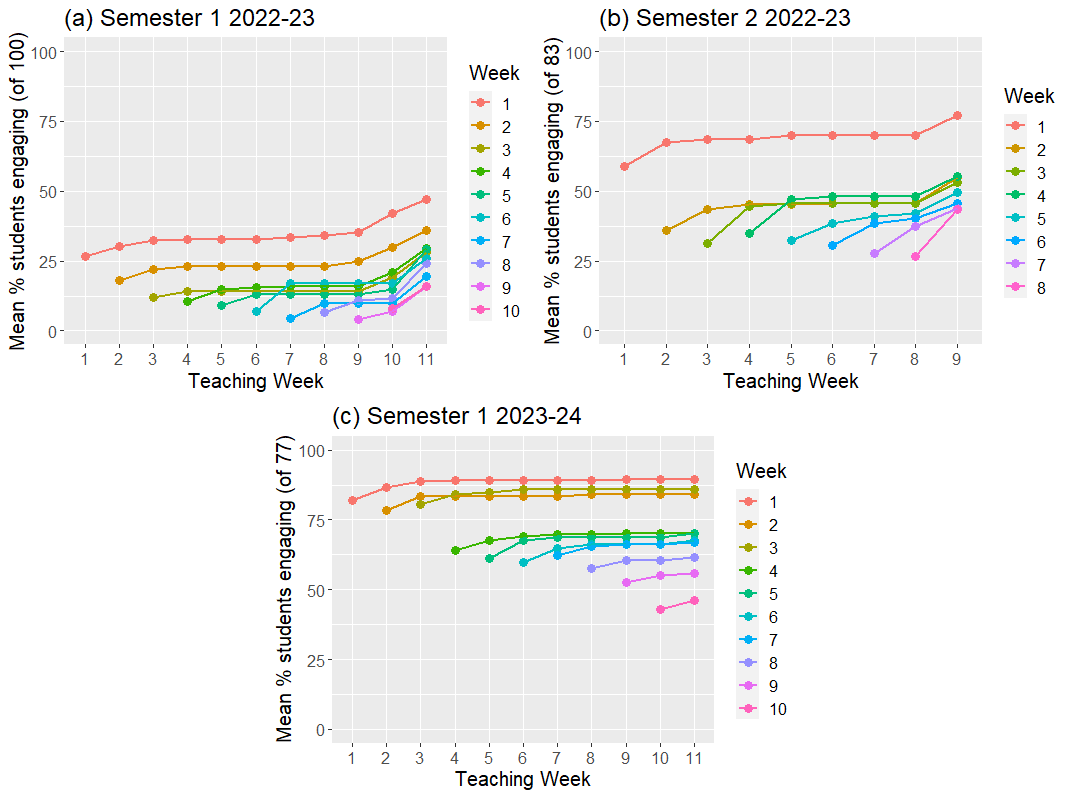}
\caption{Cumulative mean percentage of students engaging in weekly pre-lecture quizzes across each semester of interest. Figure (a) shows data for semester 1 in academic year 2022-23 before the Level Up! intervention was implemented. Figure (b) shows data for semester 2 in academic year 2022-23 after the Level Up! intervention was implemented. Figure (c) shows data for semester 1 in academic year 2023-24 after the Level Up! intervention was implemented.}\label{engagement}
\end{figure*}

The academic year of 2022-23 saw a return to campus for the majority of teaching but the issue of low engagement in pre-lecture activities persisted. This is demonstrated in Figure \ref{engagement} (a) which shows the cumulative mean percentage of students engaging in the pre-lecture quizzes for each week across an 11 week period in semester 1 of 2022-23. Each coloured line represents the cumulative average percentage of students who have engaged in the collection of quizzes associated with one week of teaching over the course of the semester (an average is shown here since each week may have a number of quizzes for students to complete). Of most interest is the level of engagement in the pre-lecture quizzes for the week in which the material is covered in class, which is represented by the first point for each weekly line in Figure \ref{engagement} (a). These figures are also presented in column 1 of Table \ref{ComparisonTable}. This figure starts off low (around 27\%) for week 1 and continues to drop for each week of material, before reaching a point around week 7 where less than 5\% of students have engaged with the pre-lecture quizzes on the week in which the material is covered in class. Another interesting feature of Figure \ref{engagement} (a) is the pattern of engagement for each collection of weekly quizzes over the semester. Although the engagement in the weekly material does increase slightly over the semester, at most only about half of the students in the class have ever engaged with the material at all on average for week 1, and after week 8 this drops to less than 25\%. Interestingly, for all weeks an increase in average engagement can be seen between weeks 10 and 11 and we believe this to be related to exam preparation as the exam takes place immediately after the course finishes. These features are clearly of huge concern to the course instructors as they indicate that not only are students not engaging with the pre-lecture quizzes `on time' to benefit from a flipped approach, the majority of students aren't engaging with the materials at all across the entire duration of the course. 


\begin{figure*}%
\centering
\includegraphics[width=\textwidth]{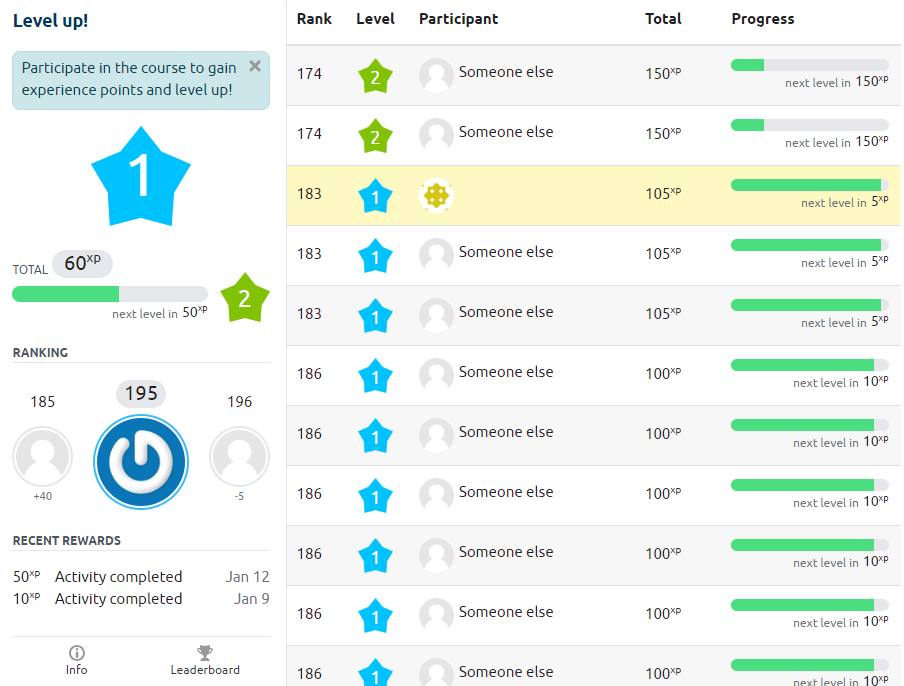}
\caption{The left-hand image shows the Level Up! block that students can see on the Moodle course page. The right-hand image shows the leaderboard view that students can see within the Level Up! block.}\label{LevelUp}
\end{figure*}

In an attempt to tackle this low engagement the decision was made to add an element of gamification to the course which would encourage students to engage in the pre-lecture activities. This was implemented through the the Level Up! Moodle plug-in\footnote{\url{https://moodle.org/plugins/block_xp}} which was introduced to the course in semester 2 of 2022-23. Level Up! allows students to gain points by participating in the course via Moodle. Points were awarded for the completion of a variety of course activities including the pre-lecture activities, tutorial questions, lab materials, and assessed quizzes and students can ``Level up" when they reach the set number of points for each level. The Level up! plug in allows the course lecturer to create rules for when points should be awarded and how many points should be awarded for activities that fall under each rule. This is entirely customisable which allows for a great deal of flexibility. The number of levels and points required to reach each level is also customisable. For this study, points were awarded based on completion of activities only and did not relate to achievement in these activities, i.e. the points system was endeavour-based and not merit-based. In an attempt to explicitly tackle engagement in the pre-lecture activities, full points were only awarded if the activity was completed before the associated lecture/tutorial/lab with half-points being awarded thereafter. Figure \ref{LevelUp} shows how this plug-in is integrated within the course Moodle page. The left-hand image shows the Level Up! block that students can see on the main course page which shows  which level they are currently on, how many points they have currently, how many they need to reach the next level, where they rank in comparison to the rest of the class and any recent points they have earned. The right-hand image shows the full leaderboard view which shows students where they rank individually on an anonymous leaderboard throughout the semester and allows them to monitor the number of points other members on the course have earned. There is also an option to view where the students' tutorial group rank in a team leaderboard, another feature which can be activated by the course lecturer and can link to course groups set-up within the Moodle page. There are additional features available in Level up! which were not implemented in this study such as conditional content which is only revealed when a certain level has been attained and drops which are invisible shortcodes worth points that can be placed anywhere on the Moodle page and ``picked up" by the user.

Level Up! was initially introduced in an attempt to increase engagement in pre-lecture activities in semester 2 of the 2022-23 academic year and, after initial success, introduced from semester 1 of the 2023-24 academic year. This allows us to make a number of comparisons which form our questions of interest: 

\begin{enumerate}
    \item How do engagement levels in pre-lecture quizzes compare after the implementation of Level Up! between the same students across two courses. \label{Q1}
    \item How do engagement levels in pre-lecture quizzes compare after the implementation of Level Up! within the same course across two academic years. \label{Q2}
\end{enumerate}

\section{Findings}\label{secfindings}
Below we will present findings which relate to the questions of interest listed in Section \ref{secmotivation}.

\subsection{Comparison within year (same students, different courses)}\label{subsecQ1}
We consider engagement levels for the same group of students across two semesters: in semester 1 Level Up! was not used, but in semester 2 it was fully implemented. Figure \ref{engagement} (b) shows the cumulative mean percentage of students engaging in the pre-lecture quizzes each week across a 9 week period in semester 2 of 2022-23. Of the 100 students who provided consent in semester 1 of 2022-23, 83 continued to study statistics in semester 2. Therefore, the data displayed in this figure relate to a subset of the same students in Figure \ref{engagement} (a) that provided consent to take part in this study across an entire academic year. It is immediately clear that overall the levels of engagement in semester 2 of 2022-23 (after Level Up! was implemented) are higher when comparing to semester 1 (before Level Up! was implemented) as shown in Figure \ref{engagement} (a). Similar to Figure \ref{engagement} (a), a drop-off in completion of pre-lecture quizzes for each week is still observed, however the baseline level of engagement for each week remains substantially higher, with the lowest mean engagement for a collection of weekly pre-lecture quizzes in the week the material is covered in-class being 25\% in Week 9. This is around the same level as for Week 1 in Figure \ref{engagement} (a), which showed the highest levels of engagement across the semester. This suggests that Level Up! has had a positive impact on engagement in pre-lecture quizzes across this student cohort. Another key feature of this figure is the clear drop-off in engagement after Week 1, which can be seen by the red line (associated with Week 1 materials) sitting considerably higher and separate to the lines representing the other weeks. This indicates that the effect of Level Up! on engagement drops more substantially from week 1 to week 2 compared to any other pairs of weeks and could be attributed to the initial excitement of the gamification and competition wearing off after week 1. Despite this, it can still be concluded that over the course of the semester the levels of engagement in the pre-lecture quizzes were higher than for semester 1. The average percentage of students engaging in the pre-lecture quizzes during the corresponding teaching week is also shown in Table \ref{ComparisonTable}. These figures correspond to the first point in each weekly line in Figures \ref{engagement} (a) and \ref{engagement} (b). It should be noted that the course structures differ between semester 1 and 2 with weeks 9 and 10 reserved for additional guest lectures and exam revision and as such these weeks were excluded since little new material was introduced.

\begin{table}[]
\begin{tabular}{|l|ccc|}
\hline
                       & \multicolumn{3}{c|}{\textbf{\begin{tabular}[c]{@{}c@{}}Mean \% of students completing pre-lecture \\ materials during teaching week\end{tabular}}}                                                                                                                                                          \\ \hline
\textbf{Teaching Week} & \multicolumn{1}{c|}{\textbf{\begin{tabular}[c]{@{}c@{}}Semester 1 2022-23\\ (n=100)\end{tabular}}} & \multicolumn{1}{c|}{\textbf{\begin{tabular}[c]{@{}c@{}}Semester 2 2022-23\\ (n=83)\end{tabular}}} & \multicolumn{1}{c|}{\textbf{\begin{tabular}[c]{@{}c@{}}Semester 1 2023-24\\ (n=77)\end{tabular}}} \\ \hline
Week 1                 & 26.7                                                                                             & 59.0                                                                                             & 81.8                                                                                            \\
Week 2                 & 18.0                                                                                             & 36.1                                                                                             & 78.4                                                                                           \\
Week 3                 & 12.0                                                                                             & 31.3                                                                                             & 80.5                                                                                            \\
Week 4                 & 10.5                                                                                             & 34.9                                                                                             & 64.0                                                                                            \\
Week 5                 & 9.0                                                                                              & 32.5                                                                                             & 61.0                                                                                            \\
Week 6                 & 7.0                                                                                              & 30.7                                                                                             & 59.7                                                                                            \\
Week 7                 & 4.3                                                                                              & 27.7                                                                                             & 62.3                                                                                            \\
Week 8                 & 6.5                                                                                              & 26.5                                                                                             & 57.8                                                                                            \\
Week 9                 & 4.0                                                                                              & N/A                                                                                                & 52.6                                                                                            \\
Week 10                & 8.0                                                                                              & N/A                                                                                                & 42.9                                                                                            \\ \hline
Level Up! Implemented  & No                                                                                                 & Yes                                                                                                & Yes                                                                                               \\ \hline
\end{tabular}
\caption{Mean percentage of students engaging pre-lecture materials during teaching week across each semester.}\label{ComparisonTable}
\end{table}

In order to confirm the increase in engagement that can be seen in semester 2 (Figure \ref{engagement} (b)) compared to semester 1 (Figure \ref{engagement} (a)), a 95\% confidence interval was calculated to compare the mean difference in the proportion of pre-lecture quizzes completed in the week the material is covered in class between semester 1 and 2 for each student who participated in the study.  The 95\% confidence interval for the mean difference in engagement between semester 1 and semester 2 was found to be $(-0.29, -0.15)$, which indicates that, on average, a higher level of engagement was found in semester 2, with this increase likely to lie somewhere between 15\% or 29\%. These results suggest that the implementation of Level Up! significantly improved the proportion of pre-lecture quizzes completed by the students.

\begin{figure*}%
\centering
\includegraphics[width=\textwidth]{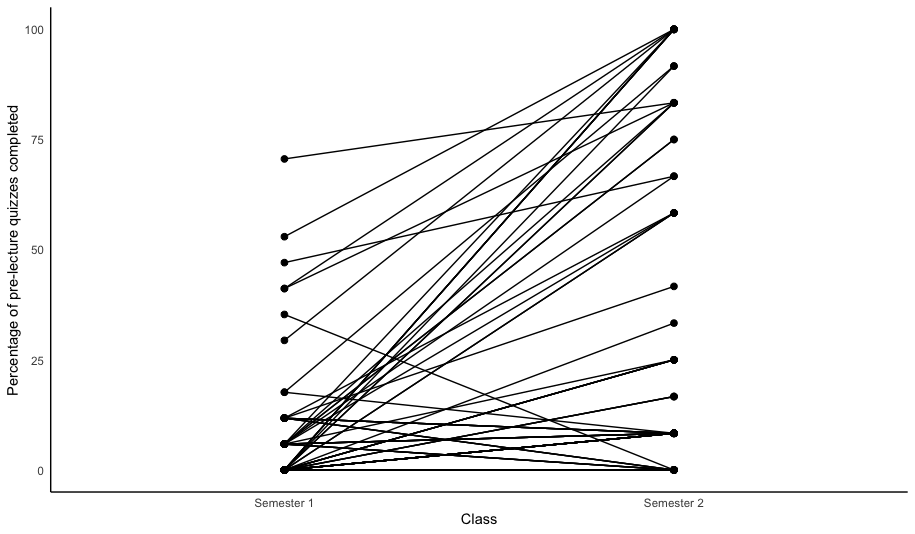}
\caption{Percentage of pre-lecture quizzes completed for the 83 students who were enrolled in both courses during 2022-23 in semester 1 (no Level Up!) and semester 2 (Level Up! implemented). There were a total of 17 pre-lecture quizzes in semester 1, and a total of 12 pre-lecture quizzes in semester 2. Not all quizzes were attempted in semester 1 by students, hence the lower number of observations.}\label{pairs}
\end{figure*}

We were also interested in whether all students were more engaged (in general) in semester 2, or if the gamification aspect was drawing out students who were not engaged in semester 1 who then go on to become very engaged in semester 2. Figure \ref{pairs} shows the average engagement levels of the 83 students who were enrolled in both courses in 2022-23 across both semesters, with each pair of points and connecting line representing one student. This figure shows some interesting features. Firstly, as we would expect, the engagement levels tend to increase for most students across both semesters, however there are some instances where engagement levels decrease. Potential reasons for the drop in engagement between semesters could be due to a lack of motivation for the concept of Level Up! and general gamification. It is also worth noting that those whose engagement dropped between semesters tended to have low engagement levels in semester 1. For students who had comparatively high levels of engagement in semester 1, we tend to see a reasonable increase in semester 2 indicating that students who were already more engaged than their peers are likely to engage further after the implementation of Level Up!. We also see a number of very sharp increases from semester 1 to semester 2 for students who were not very engaged (or not engaged at all) in semester 1, who then show some of the highest engagement levels in semester 2. For these students it seems that the gamification element drastically improved their motivation and engagement levels and is evidence that including a gamification element to a course may encourage engagement from students who would otherwise not engage.

Despite these positive results, there may be other factors which could affect the engagement levels across these semesters. Engagement levels may be impacted by the materials which are covered in each course, as these will will differ from one semester to the next, and to confound matters, the teaching team differed between semesters. However, in general, course evaluations show that students are very positive about the teaching across both semesters and rate the teaching team very highly in both courses.

It is not possible to investigate the effect of Level Up! on end-of-semester exam performance within the 2022-23 cohort since each course covers different material. However, we can assess meaningful engagement with the formative reading quizzes by monitoring grades for these quizzes. The average percentage grade for formative reading quizzes in semester 1 of 2022-23 was 68\% compared to 69\% for semester 2 of 2022-23. These percentages only include students who completed the quizzes and so we would not necessarily expect to see an increase in percentage grade after the implementation of Level Up! but it is reassuring to see that the average percentage grade is similar to the previous semester indicating that, in general, students are engaging with these quizzes in a meaningful way rather than just completing them to gain points and `game' the leaderboard.

\subsection{Comparison between years (different students, same course)}\label{subsecQ2}
 The second question of requires us to compare engagement levels for different groups of students for the same course (i.e. same semester) across two academic years. In 2022-23, Level Up! was not implemented and in 2023-24 it was. There were some small changes made to the course  materials between these years, however the course aims, intended learning outcomes and structure remained the same. Figure \ref{engagement} (c) shows the cumulative mean percentage of students engaging in the pre-lecture quizzes each week across a 11 weeks in semester 1 of the 2023-24 academic year. The data displayed in this figure relate to different students than in Figure \ref{engagement} (a) but for the same course. When comparing to Figure \ref{engagement} (a) it is again very clear to see that overall engagement levels in the pre-lecture materials were substantially higher after Level Up! was implemented. 
 
 In week 1, around 82\% of students had completed the pre-lecture quizzes, compared to only around 27\% for the same material in 2022-23. Although we still see a drop-off in engagement through the semester, the engagement levels in each week always remain substantially higher than observed the previous year. We also see a larger drop in engagement between weeks 3 and 4. The average percentage of students engaging in the pre-lecture quizzes during the corresponding teaching week is also shown in Table \ref{ComparisonTable}. These figures correspond to the first point in each weekly line in Figures \ref{engagement} (a) and \ref{engagement} (c).

In order to confirm the increase in engagement that can be seen in semester 1 of 2023-24 (Figure \ref{engagement} (c)) compared to semester 1 of 2022-23 (Figure \ref{engagement} (a)), a 95\% confidence interval  was calculated to compare the mean difference in the proportion of pre-lecture quizzes completed in the week the material is covered in class between semester 1 and 2 for each student who participated in the study.  The 95\% confidence interval for the mean difference in engagement between 2022-23 and 2023-24 is $(-0.53, -0.37)$, which indicates that, on average, a higher level of engagement was found in semester 2, with this increase likely to lie somewhere between 37\% or 53\%. These results suggest that the implementation of Level Up! significantly improved the proportion of pre-lecture quizzes completed by the students.

It was also of interest to compare achievement in this course across both cohorts to understand if the increased engagement led to better understanding of the material amongst the students. At course level, we believe the increased engagement has contributed to an increase in performance with 82\% of students receiving a grade of A, B or C in the semester 1 course of 2023-24 compared to 60\% in 2022-23. We can also assess meaningful engagement with the formative reading quizzes by monitoring grades for these quizzes. Similarly to Section \ref{subsecQ1} the average percentage grade in formative reading quizzes across the years were very similar with an average in semester 1 of 2023-24 of 64\% compared to 68\% for semester 1 of 2022-23 suggesting students were still engaging in these quizzes in a meaningful way.

As in Section \ref{subsecQ1}, it should be noted that there may be other factors which could affect the engagement levels and performance across these cohorts. Clearly the set of students in each cohort was different and so some of the changes we identified could be attributed to a cohort effect. There was also a change in the teaching team, with a change in one (of two) course lecturers. However, in general, course evaluations from each year show that students were very positive about the teaching team across both years. The changes identified in performance will also be impacted by a different exam paper, however the material covered and difficulty level of the paper was kept as consistent as possible between the two years.

Although it is less useful to compare Figure \ref{engagement} (b) and Figure \ref{engagement} (c) since these data refer to different students and different courses, it is interesting that the engagement levels in Figure \ref{engagement} (c) are substantially higher that in Figure \ref{engagement} (b), even though both courses implemented Level up! This may be explained by the observation that students are generally more engaged at the beginning of an academic year, particularly for a Level 1 course where most students are studying at University for the first time and therefore Level up! (or other teaching interventions) may be more effective if implemented at the beginning of a new academic year. In both Figures \ref{engagement} (a) and (b) we see a large drop in engagement after a certain week. Interestingly this comes after week 1 for semester 2 2022-23 and slightly later (between weeks 3 and 4) for semester 1 2023-24.

\section{Discussion}\label{secdiss}

The purpose of this study was to investigate the effects of a gamification element in a mathematical sciences course taught in flipped format. The study compared levels of engagement with pre-lecture quizzes before and after the gamification intervention was introduced. Comparisons were made across semesters (different courses, same students) and across academic years (same course, different students). Our study found that in both cases the level of engagement (measured by average percentage of students completing pre-lecture quizzes prior to the associated in-person sessions) was significantly higher when compared to a course where Level Up! was not implemented. Although engagement levels did drop over the course of each semester, they remained substantially higher for each week when compared to the previous semester or year. These results provide convincing evidence that introducing a gamification element to a course taught in flipped format increases students motivation and engagement to complete work outside of scheduled class time. 

When comparing engagement levels within students between two semesters, we found a general increase in engagement across both semesters but more interestingly found that some students with very low engagement levels in semester 1 went on the have some of the highest engagement levels in semester 2. For these students it seems that the gamification element drastically improved their motivation and engagement levels and is evidence that including a gamification element to a course may encourage engagement from students who would otherwise not engage.

It is thought that the two statistics courses considered here are the only time the students came across the Level Up! plugin or other form of gamification. It is therefore not known whether the increased engagement is due to the novelty of a leaderboard and whether gamification would be as successful if implemented widely across multiple courses.

Our results also found that this increase in engagement could lead to an increase in performance with 82\% of students receiving a grade of A, B or C in the semester 1 course of 2023-24 compared to 60\% in 2022-23. Although we cannot say for sure what led to this increase in achievement and there are many factors which could influence this (such as cohort effect, different exam paper, etc.), an increase in engagement throughout a semester is compatible with an increase in performance since students are more likely to be distributing their learning over the course of a semester. This is especially beneficial for mathematical sciences courses where learning is often thought of as linear with each topic building on the previous one and poor understanding of early material can have a detrimental effect later in the course \cite{Ireland20}.  

This study focuses on Year 1 undergraduate students who typically fall into a post-secondary education age range. The transition to university for students can be overwhelming, with changes in teaching delivery, workload and cognitive demands (\cite{Blair17}). Our results show that gamification can improve the learning experience, which may help with this transition, in line with observations by \cite{Zaric17}. This being said, \cite{Kim21} looked at the effects of gamification across three age cohorts and found that university level participants found the least significant results in terms of positive interaction when compared to secondary school and adult learners, indicating that younger and older learners may have more interest in gamified factors. Currently, gamification is only present within the Year 1 modules of the undergraduate programme, and is not implemented within other modules on the programme. There is scope to explore the effects of gamification on a more mature cohort to observe if we see similar patterns of engagement. 

This study has considered student engagement levels in terms of completing pre-lecture quizzes, though it is important to note that we see meaningful engagement instead of cases where students are simply trying to `game' the leaderboard. Our results are based solely on participation on formative reading quizzes where participation is identified through criteria being met, as this material was found to show that genuine participation took place. There are other activities in the course, such as textbook readings, which simply require a manual completion indicator and could be exposed to non-meaningful engagement. We feel that a reasonable performance on the quizzes indicates engagement with the other activities available (or indeed, of students independently finding their own resources for study).

The results presented here are for a self-selected subset of students which could bias the results as these students may be more motivated to engage compared to those who did not opt to take part. However, course instructors were able to view results for the entire cohort of students to inform their future decision making on the learning and teaching approach for this course and others. The results presented here were found to be similar to the results for the entire cohort and the general conclusions made would not change if data on the entire cohort of students were considered.  

Although studies in gamification have reflected positive outcomes in terms of increased motivation and engagement in tasks, this teaching strategy has potential downsides. Effects such as increased competition between students (\cite{Hakulinen13}) and task evaluation difficulties, where certain tasks may not be suitable for a game based element or may have ways of being marked as complete without proper engagement in the learning materials (\cite{Dominguez13}) may become issues. The design of game based elements within a course must also be considered with a balance of reward and level of work and engagement required. Tasks which are overly complex can lead to a lack of engagement from learners (\cite{Dong12}). There is also potential for demotivation through the full gamification process, particulaly if students find themselves placed low on the points leaderboard. The competency levels of students may lead to negative consequences, including frustration and loss of self confidence (\cite{Alomari19}). A study by \cite{Ding18} showed that some students in a gamified learning activity may require more time to obtain the necessary understanding to progress through learning process. 

Engagement in gamification not only requires active participation from students, but time and attention from staff to implement the system effectively. From our experience using Level Up!, it takes time to set up the system. Several considerations have to be made at the set-up stage, including the allocation of experience points to certain tasks, and allocating points in such a way as to reflect the level of engagement required from the student. Levels were attainable for students which were split by weekly tasks, which requires manual intervention to tally the available points over a given week. For example, students could receive half points for completing certain tasks after a deadline which requires manual intervention from the course lecturer to reset the points awarded to a different total after each deadline. Though our study explored gamification using Level Up!, there are a host of other plug-ins which can be adapted in virtual learning environments such as Quizventure\footnote{\url{https://github.com/xow/moodle-mod_quizgame}} and Block Game\footnote{\url{https://moodle.org/plugins/block_game}}. \cite{Heilbrunn17} provide a detailed review of tools for gamification analytics. 

Finally, we highlight the importance of using learning analytics, where possible, to access data from students on a course when measuring the success of a new teaching method or intervention. If the  instructors of the two courses under consideration here had relied on feedback from students to inform teaching strategy - which was consistently positive even before the implementation of Level Up! - the true extent of the lack of engagement in the pre-lecture materials may never have been identified. 

\section{Conclusion}\label{secconc}
The results we present here are in line with previous research (for example \cite{Lo17}, \cite{Buckley16}, \cite{Dahlstrom12}) and show a stark increase in student engagement with pre-lecture activities on a course taught in flipped mode when gamification strategies were employed. We found no evidence that the increased engagement was due to students `gaming' the leaderboard.

A variety in teaching and learning approaches is surely the key to motivating students across their courses, with gamification just one strategy in a teacher's toolbox. Setting up a gamified environment takes time, and our positive results may be the consequence of the novelty of Level Up!. The implementation of gamification strategies should therefore be carefully considered, not just course by course but across a student's programme of study to ensure a diversity of learning approaches.  

\section{Acknowledgements}
All data supporting this study are provided as supplementary information accompanying this paper at \url{https://doi.org/10.5525/gla.researchdata.1656}.

\section{References}


\end{document}